\documentclass[12pt]{oxarticle}
\pdfoutput=1
\usepackage{color,graphicx, float, array, xspace, amscd, amsmath, amsthm, amssymb, latexsym,  mathtools, caption}
\usepackage{bbm, setspace}
\usepackage[sort&compress, comma, square, numbers]{natbib}
\usepackage{xcolor}
\usepackage{navigator}

\let\SS=\S 
\let\nordico=\o 
\renewcommand{\a}{\alpha}
\renewcommand{\b}{\beta}
\newcommand{\g}{\gamma}
\renewcommand{\d}{\delta}
\newcommand{\e}{\epsilon}
\newcommand{\z}{\zeta}

\renewcommand{\k}{\kappa}
\renewcommand{\l}{\lambda}

\newcommand{\p}{\pi}

\newcommand{\s}{\sigma}\renewcommand{\S}{\sum}

\renewcommand{\o}{\omega}

\DeclareFontFamily{OT1}{pzc}{}
\DeclareFontShape{OT1}{pzc}{m}{it}{<-> s * [1.200] pzcmi7t}{}
\DeclareMathAlphabet{\mathpzc}{OT1}{pzc}{m}{it}

\newcommand{\cG}{\mathcal{G}}

\newcommand{\one}{\mathbbm{1}}

\newcommand{\IC}{\mathbb{C}}

\newcommand{\IP}{\mathbb{P}}
\newcommand{\IQ}{\mathbb{Q}}

\newcommand{\IZ}{\mathbb{Z}}

\font\eightrm=cmr8 at 8pt
\font\csc=cmcsc10

\DeclareFontFamily{U}{wncy}{}
\DeclareFontShape{U}{wncy}{m}{n}{<->wncyr10}{}
\DeclareSymbolFont{mcy}{U}{wncy}{m}{n}
\DeclareMathSymbol{\sha}{\mathord}{mcy}{"58}

\newcommand{\varstr}[2]{\vrule height #1 depth #2 width0pt}

\newcommand{\place}[3]{\vbox to0pt{\kern-\parskip\kern-7pt
                             \kern-#2truein\hbox{\kern#1truein #3}
                             \vss}\nointerlineskip}
                             
\newcommand{\capt}[3]{\parbox{#1}{\renewcommand{\baselinestretch}{1.0}
                                                           \caption{\label{#2}\small\it #3}}}

\newcommand{\beq}{\begin{equation}}
\newcommand{\eeq}{\end{equation}}
\newcommand{\beqnn}{\begin{equation*}}
\newcommand{\eeqnn}{\end{equation*}}

\newcommand{\tref}[1]{Table~\ref{#1}}
\newcommand{\sref}[1]{\SS\ref{#1}}

\newcommand{\defineas}{\buildrel\rm def\over =}

\newcommand{\del}{\partial}

\newcommand{\+}{\hphantom{-}}

\newcommand{\cicy}[2]{\begin{matrix} #1\end{matrix}\!\left[\begin{matrix}#2 \end{matrix}\right]}

\newcommand{\quotient}[1]{_{\hskip-2pt\lower1pt\hbox{$/$}\lower2pt\hbox{\hskip-1pt$#1$}}}
\renewcommand{\baselinestretch}{1.1}

\numberwithin{equation}{section}
\proofmodefalse
\begin{document}
\pagestyle{empty}
\begin{center}
\null\vskip0.3in
{\Huge {Highly Symmetric Quintic Quotients}\\[0.59in]}
{\csc Philip Candelas$^1$ and Challenger Mishra$^2$\\[1.3cm]}
{\it $^1$Mathematical Institute\hphantom{$^1$}\\
University of Oxford\\
Radcliffe Observatory Quarter\\ 
Woodstock Road,
Oxford OX2 6GG, UK\\[4ex]
$^2$Rudolf Peierls Centre for Theoretical Physics\hphantom{$^3$}\\
University of Oxford\\
1 Keble Road, 
Oxford OX1 3NP, UK\\
}

\vfill
{\bf Abstract\\[2ex]}
\parbox{6.0in}{\setlength{\baselineskip}{14pt}
The quintic family must be the most studied family of Calabi-Yau threefolds. Particularly symmetric members of this family are known to admit quotients by freely acting symmetries isomorphic to $\IZ_5{\times}\IZ_5$. The corresponding quotient manifolds may themselves be symmetric. That is, they may admit symmetries that descend from the symmetries that the manifold enjoys before the quotient is taken. The formalism for identifying these symmetries was given a long time ago by Witten and instances of these symmetric quotients were given also, for the family $\IP^7[2,2,2,2]$, by Goodman and Witten. We rework this calculation here, with the benefit of computer assistance, and provide a complete classification. Our motivation is largely to develop methods that apply also to the analysis of quotients of other CICY manifolds, whose symmetries have been classified recently. For the $\IZ_5{\times}\IZ_5$ quotients of the quintic family, our list contains families of smooth manifolds with symmetry $\IZ_4$,
$\text{Dic}_{3}$ and $\text{Dic}_{5}$,
families of singular manifolds with four conifold points, with symmetry $\IZ_6$ and $\IQ_8$, and rigid manifolds, each with at least a curve of singularities, and symmetry $\IZ_{10}$. We intend to return to the computation of the symmetries of the quotients of other CICYs elsewhere. 
}
\end{center}
\newpage
\begingroup
\baselineskip=14pt
\tableofcontents
\endgroup
\newpage
\setcounter{page}{1}
\pagestyle{plain}
\section{Introduction}
\vskip-10pt
Many Calabi-Yau manifolds are known, but until fairly recently very few were known that have small Hodge numbers\footnote{By Hodge numbers $(h^{1,1},h^{2,1})$ being small, somewhat loosely we mean $h^{1,1}+h^{2,1}\le 20$.}. Finding such manifolds seems to require taking quotients, usually freely acting quotients of simply connected manifolds. This process of taking quotients then does double duty: on the one hand, the Hodge numbers are reduced and, on the other, the quotient manifold acquires a nontrivial fundamental group that permits flux breaking on the $E_8{\times}E_8$ gauge group of the heterotic string, to a group that is of greater interest for the purpose of  constructing a realistic heterotic string vacuum. 

This much is an old story. The more recent element is that it has been possible to find freely acting quotients in a systematic way, albeit for a special class of Calabi-Yau manifolds, the so called CICYs (complete intersection Calabi-Yau manifolds). An important work in this direction was the systematic classification by V\!.~Braun \cite{Braun:2010vc} of all freely acting quotients of manifolds on the CICY list. 

Further work in \cite{Candelas:2008wb, Candelas:2010ve, candelas2016hodge, constantin2017hodge} calculated the Hodge numbers for each of these quotients, and was summarised in \cite{candelas2016calabi}. Not all of the quotient manifolds seem  interesting, but some are. Among these is the familiar example of the $\IZ_5{\times}\IZ_5$ quotient of the quintic: 
\beq\label{otherquotients}
X={\IP^4[5]}\quotient{\IZ_5{\times}\IZ_5},~~~(h^{1,1}(X), h^{2,1}(X))=(1,5).\notag
\eeq
A number of other interesting quotients have a somewhat similar structure being quotients by a group $\IZ_{p^{n}}{\times}\IZ_{p^{n}}$ for a prime power $p^n$. These include:
\beq\label{Heisenberg}
\IP^7[2,2,2,2]\quotient{\IZ_8{\times}\IZ_8},~~~~~~~~\cicy{\IP^2\\ \IP^2}
{ ~3 \!\!\!\!\\
  ~3\!\!\!\! & \\}\quotient{\IZ_3{\times}\IZ_3}
\eeq
and the $\IZ_7\times\IZ_7$ quotient of the R{\nordico}dland manifold \cite{Rodland:1998pm}. The first manifold in \eqref{Heisenberg} denotes the resolved quotients of resolutions of singular complete intersections of four quadrics in $\IP^7$ \cite{freitag2011siegel}. A question that naturally arises is what symmetries do these quotient manifolds have? Such remnant symmetries may manifest themselves as global symmetries of the corresponding low energy theory and are thus potentially phenomenologically important. For the quintic quotient the general formalism for addressing this question was given by Witten \cite{Witten:1985xc} (in perhaps the second paper to be written on string compactification on Calabi-Yau manifolds). A number of the possible symmetries, or rather the symmetric quintic quotients, are given in that paper and further cases were provided by Goodman and Witten in~\cite{goodman1986global}. A number of quintics symmetric under finite simple groups were also constructed in \cite{luhn2008quintics} employing various group and representation theory techniques and data from the ATLAS of finite groups \cite{conway1985atlas}.

Our interest is principally with the other quotients listed above  \eqref{otherquotients},  but any attempt at a complete treatment of these cases requires a certain amount of machine calculation. Developing the routines to analyse these quotients has brought us back to the quintic for which we reproduce the results of the classic papers together with a number of extensions. We intend to return to a parallel treatment of the quotients \eqref{Heisenberg} in a future work.  In this paper, we focus on the quintic family $\IP^4[5]/ \IZ_5{\times}\IZ_5$, with a view to constructing sub-families  with non-trivial symmetries. 
\vskip1cm
\section{Preliminaries}\label{stateoftheart}
\vskip-10pt
The quintic is defined as the zero-locus of a homogeneous polynomial of degree 5 with variables in $\IP^4$, denoted by $\IP^4[5]$. We are concerned with the case that the manifold admits freely acting symmetries by the following two order 5 generators:
\vskip-0.4cm
\beq\label{eq:quintic_repQ}
S: x_i\rightarrow \zeta^i x_i
~~~\text{and}~~~ 
T: x_i\rightarrow x_{i-1}
\eeq
where the $x_i~(i\in \IZ_5)$ are the homogeneous co-ordinates of $\IP^4$ and $\zeta$ is a non-trivial fifth root of unity. Note that $\langle S,\ T\rangle\!=\IZ_5{\times}\IZ_5$. The most generic quintic, invariant under both $S$ and $T$ can be written as the zero locus of the following 5 parameter\footnote{There are six coefficients $c_a$, but an overall scale is irrelevant, since it does not change the zero locus of the polynomial.} family of degree 5 polynomials:
\vskip-1cm
\begin{align}\label{Z5Z5quinticQ}
\hskip4.25cm&\hskip-4.25cm x\in\IP^4[5]\quotient{\IZ_5{\times}\IZ_5} ~~~\text{satisfies}~~~
p(x)\;\defineas\;\sum_{a=1}^{6} c_{a}{\bf J}_a\;=\;0\notag\\[5pt]
\hidewidth{\text{where}~~~}
{\bf J}_1=&~{\prod}_i\, x_i \notag\\[4pt]
{\bf J}_2=&~{\frac{1}{5}}\,{\sum}_i\, x_{i-1}^2\,x_i\,x_{i+1}^2\notag\\[4pt]
{\bf J}_3=&~{\frac{1}{5}}\,{\sum}_i\, x_{i-2}^2\,x_i\,x_{i+2}^2\\[4pt]
{\bf J}_4=&~{\frac{1}{5}}\,{\sum}_i\, x_{i-2}\,x_i^3\,x_{i+2}\notag\\[4pt]
{\bf J}_5=&~{\frac{1}{5}}\,{\sum}_i\, x_{i-1}\,x_i^3\,x_{i+1}\notag\\[4pt]
{\bf J}_6=&~\frac{1}{5}\,{\sum}_{i}\, x_i^5 \notag
\end{align}
The quintic and its quotient by the above freely acting $\IZ_5\times\IZ_5$ symmetry are distinct families of manifolds evidenced by different Hodge numbers. The Hodge pairs for the quintic, and its quotients by freely acting $\IZ_5$ and $\IZ_5{\times}\IZ_5$ symmetries are (1,101), (1,21) and (1,5) respectively.

The generic quintic is a linear combination of 126 distinct monomials. One can think of \eqref{Z5Z5quinticQ} as defining the sub-family of quintics that is left invariant by the symmetry actions~\eqref{eq:quintic_repQ}. It is natural to ponder whether one could obtain quintic families with even larger symmetries upon further restriction of the defining polynomial. 

Before we embark on finding such families it is worth noting that \eqref{Z5Z5quinticQ} admits a $\IZ_2$ symmetry without any further restriction of $p(x)$. This global discrete symmetry is: 
\beq\label{eq:quintic_new_symm1}
\pi^{(2)}: ~~~x_i \;\rightarrow\; x_{4-i}~.
\eeq
This operation reverses the order of the coordinates and we see from \eqref{Z5Z5quinticQ} that this preserves each 
${\bf J}_a$. Symmetry groups of any sub-family of the quintic quotient must then contain this $\IZ_2$ as a subgroup. This provides an useful restriction in deciding which symmetry groups are possible. We will find these symmetry groups in the following, but we summarize here the main results of this work in \tref{summary}. 
\vskip0.2cm
\begin{table}[H]
\begin{center}
\begin{tabular}{| c || c | c | c | c | c |}
\hline\varstr{16pt}{10pt}Symmetry Group  & \# parameters & \# sub-classes & ~smooth/singular~ 
& ~~Reference~~ \\
\hline\hline
\varstr{16pt}{10pt} $\IZ_4$ & 3 & 15 & smooth & \tref{TZ4}\\
 \hline
\varstr{16pt}{10pt} $\IZ_6$ & 1 & 20 & 4 conifold points &  \tref{TZ6}\\
 \hline
\varstr{16pt}{10pt} $\IQ_8$ & 1 & 15 & 4 conifold points & \tref{TQ8}\\
 \hline
\varstr{16pt}{10pt} $\IZ_{10}$ & 0 & 24 & singular curve(s) &\tref{TZ10}\\
 \hline
\varstr{16pt}{10pt} $\text{Dic}_{3}$ & 1 & 10 & smooth & \tref{TDic3}\\
  \hline
\varstr{16pt}{10pt} $\text{Dic}_{5}$ & 1 & 6 & smooth & \tref{TDic5}\\
  \hline
 \end{tabular}
 \vskip 0.5cm
\capt{6.15in}{summary}{An overview of the results in this paper. The quintic quotient has sub-families with the above listed symmetries.  The table lists also the number of distinct sub-families corresponding to each symmetry group. Each of the groups listed above contains the $\IZ_2$ defined in \eqref{eq:quintic_new_symm1} as a subgroup. A detailed description of the families is given in the tables in \sref{results_quintic_quotients}.}
\end{center}
 \vskip -1.0cm
\end{table}
The table lists the various symmetry groups and the corresponding number of families of the quintic quotient \eqref{Z5Z5quinticQ}. In summary, we have found one parameter families of quintic quotients with remnant  $\IZ_6$, $\IQ_8$, $\text{Dic}_{3}$ and $\text{Dic}_{5}$ symmetries\footnote{Here $\IQ_8$ is the quarternion group. The dicyclic group Dic$_n$ is a non-Abelian group of order $4n$ with the presentation 
$$ \langle~ a,\,b \mid a^{2n} \;=\; 1,\ b^2 = a^n,\ b^{-1}ab = a^{-1}~\rangle~.$$ 
It can be expressed as an extension of $\IZ_2$ by $\IZ_{2n}$ as demonstrated by the following short exact sequence: $1 \to \IZ_{2n} \to \mbox{Dic}_n \to \IZ_2 \to 1$.} and several three parameter families with $\IZ_4$ symmetries. In addition, we have also found $24$ isolated quintic quotients with $\IZ_{10}$ symmetries. In the following we outline our methods. The coordinate representations and covariants of these symmetries are tabulated in \sref{results_quintic_quotients}.

\subsection{Symmetries of quintic quotients from automorphisms of $\IP^4$}\label{SymmQuinQuo}
\vskip-10pt
The linear symmetries of the quintic quotients derive from linear automorphisms of~$\IP^4$. It is shown in \cite{Witten:1985xc} that any such symmetry action $\pi\in \text{PGL}(5,\IC)$, that descends to the quotient by a freely acting group $G_\text{f}$, must satisfy the normalizer condition:
\beq
\begin{aligned}\notag
\pi\ g\ \pi^{-1} \in G_\text{f}~;~~~\forall g \in G_\text{f}
\end{aligned}
\eeq
This ensures that points that lie on the quintic, transform sensibly under the combined action of any $g\in G_\text{f}$ and any new symmetry $\pi$. For $G_\text{f}\cong\IZ_5\times\IZ_5$ generated by S and T defined in \eqref{eq:quintic_repQ}, the above normalizer condition translates to: 
\beq\label{QuinticNormalizerCond1}
\begin{aligned}
\pi\ S\ \pi^{-1}&\;=\; S^\alpha\  T^\beta \\[3pt]
\pi\ T\ \pi^{-1}&\;=\; S^\gamma\  T^\delta
\end{aligned}
\eeq
where $\alpha, \beta, \gamma, \delta \in \IZ_5$. One additionally notes from \eqref{eq:quintic_repQ} that $S\ T\ S^{-1}\ T^{-1}{\;=\;}\zeta$. Combined with the equation above, this implies that $\alpha\delta-\beta\gamma\equiv1~\text{mod}~5$, or alternatively,
\beq\notag
\sigma\defineas\left(\begin{matrix}
 \alpha & \beta \\
\gamma & \delta 
\end{matrix}\right) ~\in ~\text{SL}(2, \IZ_5)
\eeq
Reproducing the argument from \cite{Witten:1985xc}, we note that there is a one-to-one correspondence between linear projective actions $\pi$ and matrices $\sigma\in\text{SL}(2,\IZ_5)$ in the following sense. If two projective linear actions $\pi_1$ and $\pi_2$ satisfy \eqref{QuinticNormalizerCond1}, for a given $\sigma$, one can show that 
$\pi_1^{-1} \pi_2$ lies in the centralizer of $G_\text{f}$ in PGL(5,$\IC$),
which turns out to be $G_\text{f}$ itself. Thus given $x\in \IP^4[5]/G_\text{f}$, $\pi_1 x{\;=\;}\pi_2 x$. This establishes that the two projective matrices $\pi_1$ and $\pi_2$ have equivalent actions on the points of the quotient manifold $\IP^4[5]/G_\text{f}$. Thus any symmetry group of the quintic quotient is isomorphic to a subgroup of $\text{SL}(2, \IZ_5)$. 

There are a total of 75 non-trivial subgroups $R{\,\subset\,}\text{SL}(2, \IZ_5)$. These are listed in \tref{SL_2_5_Subgroups}. There is a unique $\IZ_2$ subgroup generated by $\sigma=-\one_2$. The linear action 
$\pi$, corresponding to this, is the global $\IZ_2$ symmetry $\pi^{(2)}$ defined in \eqref{eq:quintic_new_symm1}, by which we mean that this symmetry acts trivially on all points in the complex structure moduli space of the quintic quotient.  This has the consequence that there can be no quotient in the family $\IP^4[5]/{\IZ_5{\times}\IZ_5}$, with a full symmetry group that does not contain this $\IZ_2$ as a subgroup. This rules out the cases that $R=\IZ_3~\text{or}~\IZ_5$. In other words: no quotient manifold can have symmetry that is precisely $\IZ_3$, for example, since any such manifold has symmetry that is at least $\IZ_6$.
\newcommand{\skp}{\hskip5pt}
\begin{table}[H]
\vspace*{0pt}
\begin{center}
\begin{tabular}{| c || >{\skp} c <{\skp}  | >{\skp} c <{\skp} | >{\skp} c <{\skp} | >{\skp} c <{\skp} 
| >{\skp} c <{\skp} | >{\skp} c <{\skp} |>{\skp} c <{\skp} 
| >{\hskip1pt} c <{\hskip1pt} | >{\hskip1pt} c <{\hskip1pt} 
|>{\hskip-2pt}c<{\hskip-2pt}|>{\hskip-2pt}c<{\hskip-2pt}|}
\hline
\varstr{16pt}{10pt} {~$R$~} &
{$\IZ_2$} &
{$\IZ_3$} &
{$\IZ_4$}  &
{$\IZ_5$}  &
{$\IZ_6$} &
{$\IQ_8$} &
{$\IZ_{10}\!$} &
{$\text{Dic}_3$} &
{$\text{Dic}_5$} &
\small{$\text{SL}(2,\IZ_3)$} &
\small{$\text{SL}(2,\IZ_5)$}\\ 
\hline\hline
\varstr{15pt}{8pt} \# & 1 & 10 & 15 & 6 & 10 & 5 & 6 & 10 & 6 & 5 & 1 \\
\hline
\varstr{15pt}{8pt} $|\,R\,|$ & 2 & 3 & 4 & 5 & 6 & 8 & 10 & 12 & 20 & 24 & 120 \\
\hline
 \end{tabular}
\vskip 0.3cm
\capt{6.2in}{SL_2_5_Subgroups}{The non-trivial subgroups $R\subset\text{SL}(2,\IZ_5)$ together with the number of distinct homomorphic copies, obtained using GAP \cite{GAP4}. The last row records the order of the subgroups.}
 \end{center}
 \end{table}
Therefore in order to find families of quintic quotients with a given residual symmetry group 
$R{\;\subset\;}\text{SL}(2,\IZ_5)$, one should solve the normaliser condition \eqref{QuinticNormalizerCond1} for all matrices $\sigma\in R$. In practice however, solving only for the generators of $R$ suffices.
The normalizer condition ensures that an additional symmetry action $\pi$ sensibly transforms the elements of the freely acting $\IZ_5{\;\times\;}\IZ_5$, as it acts on $\IP^4$. To ensure that $\pi$ will be a symmetry of this quotient manifold,  we should check that $p(\pi x){\,=\,} \l(\pi)\,p(x)$,  for some non-zero multiplier $\l(\pi)$ and a non-trivial sub-space of the six dimensional vector space $\IC^6$ spanned by the coefficients $c_a$ of \eqref{Z5Z5quinticQ}. For the global $\IZ_2$ symmetry $\pi^{(2)}$ in \eqref{eq:quintic_new_symm1}, this is the entire $\IC^6$, since 
$p\left(\pi^{(2)}x\right){\;=~}p(x)~\forall x$. 

The action of $\pi$ on $x$ induces a linear map $A_{\pi}$ on the vector of coefficients $c$.
The eigenspaces of this linear map with non-zero eigenvalues therefore define sub-classes of the quintic quotient with an additional symmetry generated by $\pi$. The examples below will demonstrate the whole program of obtaining various quintic quotient families with additional symmetries.

\subsection{A family of $\IZ_4$ quintics}\label{Z4_example}
\vskip-10pt
Consider the subgroup $\IZ_4$ of $\text{SL}(2, \IZ_5)$ generated by 
\beq
\sigma_1{\,=\,}\left(
\begin{matrix}
2 & 0 \\[2pt]
1 & 3 \\
\end{matrix}\right)~. 
\notag\eeq
The normalizer condition \eqref{QuinticNormalizerCond1} reads:
\vskip-20pt
\beq
\begin{aligned}
\pi_1\, S\, \pi_1^{-1}&= S^2\, T^0\\
\pi_1\, T\, \pi_1^{-1}&= S^1\, T^3
\end{aligned}
\label{pi1eq}\eeq
where $S$ and $T$ are as defined in \eqref{eq:quintic_repQ}. Solving this set of equations we obtain the unique action $\pi_1$ below on $\IP^4$, up to equivalence of group actions over the quintic quotient. 
\beq
\pi_1\;=\; \left(
\begin{array}{lcccr}
 1~ & 0 & 0 & 0 & ~0 \\
 0 & 0 & \zeta ^4 & 0 & 0 \\
 0 & 0 & 0 & 0 & 1 \\
 0 & \zeta ^3 & 0 & 0 & 0 \\
 0 & 0 & 0 & \zeta ^3 & 0 \\
\end{array}
\right)~.
\label{pi1action}\eeq
\newpage
It is straightforward to check that $\pi_1^4{\;\propto\;}\one_5$. This group action $\pi_1$, induces a linear action on the vector of coefficients $c_a$ via the map $c\rightarrow c~A_{\pi_1}$, with 
\beq
~~~A_{\pi_1}=\left(
\begin{array}{llcccr}
 1~ & 0~ & 0 & 0 & 0 & \,0 \\
 0 & 0 & \zeta ^4 & 0 & 0 & 0 \\
 0 & \zeta  & 0 & 0& 0 & 0 \\
 0 & 0 & 0 & 0 & \zeta ^3 & 0 \\
 0 & 0 & 0 & \zeta ^2 & 0 & 0 \\
 0 & 0 & 0 & 0 & 0 & 1 \\
\end{array}
\right)~.
\label{Api1}\eeq
Thus $\pi_1$ preserves a sub-class of the quintic quotient if $c$ is a left eigenvector of $A_{\pi_1}$. The eigenvalues of this matrix are $-1$ and $1$. The corresponding eigenspaces, denoted by $\text{E}_{-}(\pi_1)$ and $\text{E}_{+}(\pi_1)$, have dimension 2 and 4 respectively, and are generated as follows: 
\begin{align*}
\text{E}_{+}(\pi_1) \;&= \; \left\langle\ \left(1,0,0,0,0,0\right),~\left(0,0,0,\zeta^2,1,0\right),~\left(0,\zeta,1,0,0,0,0\right),~\left(0,0,0,0,0,1\right)\ \right\rangle~, \\[3pt]
\text{E}_{-}(\pi_1) \;&=\; \left\langle\ \left(0,0,0,-\zeta ^2,1,0\right),~\left(0,-\zeta,1,0,0,0\right)\ \right\rangle~.
\end{align*}

The $\IZ_4$ covariant combinations of $\{\mathbf{J}_a\}$ corresponding to $\text{E}_{+}(\pi_1)$ are
\beq
 \mathbf{J_1},~\zeta^2\mathbf{J_4}+ \mathbf{J_5},~\zeta\mathbf{J_2}+ \mathbf{J_3}, ~\mathbf{J_6}
\notag\eeq
while, for $\text{E}_{-}(\pi_1)$, they are
\beq
-\zeta^2 \mathbf{J_4}+\mathbf{J_5}, ~ -\zeta \mathbf{J_2}+\mathbf{J_3},
\notag\eeq
The eigenspace $\text{E}_{+}$ thus corresponds to a three-parameter family with a $\IZ_4$ symmetry, while the eigenspace $\text{E}_{-}$ corresponds to a further one-parameter family of manifolds, which also have $\IZ_4$ symmetry. However, this latter family enjoys a higher symmetry $\IQ_8{\,\subset\,}\IZ_4$, as we will see presently.
\subsection{A family of $\IQ_8$ quintics}\label{Q8_example}
\vskip-10pt
Consider the group generated by the matrices $\s_1$, as above, and
\beq
\sigma_2 \;=\; \left(\begin{matrix}
 1 & 1 \\
 3 & 4 \\
\end{matrix}\right)~;
\notag\eeq
as an abstract group $\langle\s_1,\,\s_2\rangle {\;\cong\;} \IQ_8$.

This $\IQ_8$ contains the $\IZ_4$ group of the previous example. To obtain all the manifolds that exhibit this group of symmetries, we should solve the normaliser condition \eqref{QuinticNormalizerCond1} for coordinate actions $\p_1$ and $\p_2$. We have already the solution $\pi_1$ corresponding to $\s_1$ and we find the co-ordinate action corresponding to $\s_2$, to be given by
\beqnn
\pi_2 \;=\; {\left(
\begin{array}{ccccc}
 1 & \zeta  & \zeta  & 1 & \zeta ^3 \\
 \zeta  & \zeta  & 1 & \zeta ^3 & 1 \\
 \zeta ^3 & \zeta ^2 & 1 & \zeta ^2 & \zeta ^3 \\
 \zeta  & \zeta ^4 & \zeta  & \zeta ^2 & \zeta ^2 \\
 1 & \zeta ^2 & \zeta ^3 & \zeta ^3 & \zeta ^2 \\
\end{array}
\right)}\raisebox{-30pt}{.}
\eeqnn
The linear actions on the coefficients $c_a$ induced by $\pi_1$ and $\pi_2$ are then found to be $A_{\pi_1}$ and $A_{\pi_2}$, where $A_{\pi_1}$ is as in \eqref{Api1}, and 
\beqnn
A_{\pi_2}\;=\;\left(\!\!\!\!
\begin{array}{cccccc}
 -1~~ & 25 \zeta ^3 & 25 \zeta ^2 & -25 \zeta  & -25 \zeta ^4 & 5 \\[2pt]
 \+20 \zeta ^2 & 25 \left(\zeta\; +\zeta ^4\right) & 25 \left(\zeta+\zeta^2\right) & 0 & 0 & 5 \zeta ^2 \\[2pt]
 \+20 \zeta ^3 & 25 \left(\zeta ^3+\zeta^4\right) & 25 \left(\zeta +\zeta ^4\right) & 0 & 0 & 5 \zeta ^3 \\[2pt]
 -30 \zeta ^4 & 0 & 0 & -25 \left(\zeta^2 +\zeta^3\right) & -25 \left(\zeta ^2+\zeta^4\right) & 5 \zeta ^4 \\[2pt]
 -30 \zeta  & 0 & 0 & -25 \left(\zeta\; +\zeta^3 \right) & -25 \left(\zeta ^2 + \zeta^3\right) & 5 \zeta  \\[2pt]
 120 & 150 \zeta ^3 & 150 \zeta ^2 & 100 \zeta  & 100 \zeta ^4 & 5 \\
\end{array}\!\!
\right)\raisebox{-40pt}{.}\\[5pt]
\eeqnn

The eigenvalues of $A_{\pi_1}$ are, as we have noted already, $1$ and $-1$ and have multiplicities 4 and 2 respectively. The matrix $A_{\pi_2}$ has eigenvalues $+25 \left(2 \zeta ^3+2 \zeta ^2+1\right)$ and 
$-25 \left(2 \zeta ^3+2 \zeta ^2+1\right)$, with multiplicities 2 and 4 respectively. The corresponding four eigenspaces are now denoted by $E_{\pm}(\pi_1)$ and $E_{\pm}(\pi_2)$. To obtain quintic quotient sub-classes that are covariant under both the actions of ${\pi_1}$ and ${\pi_2}$, we must find those vectors $c_a$ that lie at the intersection of eigenspaces of $A_{\pi_1}$ and $A_{\pi_2}$. It is straightforward to derive the following:
\begin{align}
E_{-}(\pi_1)\cap  E_{+}(\pi_2) \;&=\; \emptyset\notag \\[3pt]
E_{-}(\pi_1)\cap  E_{-}(\pi_2) \;&=\; \left\langle(0,0,0,-\zeta^2,1,0),~(0,-\zeta,1,0,0,0)\right\rangle
\notag\end{align}
The covariants corresponding to the space $E_{-}(\pi_1)\cap  E_{-}(\pi_2)$ are thus $-\zeta^2 \mathbf{J_4}+\mathbf{J_5}$ and $-\zeta \mathbf{J_2}+\mathbf{J_3}$, which we saw in the previous example, but now note that they enjoy a $\IQ_8$ symmetry. Finally we observe that the spaces 
\beq
E_{+}(\pi_1)\cap  E_{+}(\pi_2)~~~\text{and}~~~E_{+}(\pi_1)\cap  E_{-}(\pi_2) 
\notag\eeq
both have dimension 2 and the corresponding covariants yield two further one-parameter families of manifolds with residual $\IQ_8$ symmetry. The three families that we have found in this example are listed in the first row of \tref{TQ8}. 

In principle, we should perform an analysis analogous to that of \sref{Z4_example}, above, for all the 15 $\IZ_4$ subgroups of $\text{SL}(2,\IZ_5)$, however, by appealing to the Galois group action $\z\rightarrow \z^\k$, for
$\k{\,\in\,}\{1,2,3,4\}$ we are able to reduce the number of cases that have to be analysed, as we explain in the following section. The upshot is that we find 15 one-parameter families and 15 three-parameter families of manifolds with $\IZ_4$ symmetry. Each of the one-parameter families coincides with a one-parameter family of manifolds with a 
$\IQ_8$ symmetry. These one-parameter families are grouped together with the other $\IQ_8$ families in \tref{TQ8}. The three-parameter families however do not possess a symmetry larger than $\IZ_4$ and these are listed in \tref{TZ4}. 

All the quintic quotients with remnant symmetries $\IZ_{6}$, Dic$_3$ or Dic$_5$ turn out to be one-parameter families. The quintic quotients with precisely $\IZ_{10}$ symmetry have no free parameters. Without repeating the previous analysis, for all the cases, we record the results in the tables of~\sref{results_quintic_quotients}. 

The examples above illustrate the method of computing quintic quotient families with any symmetry group $R\subset\text{SL}(2,\IZ_5)$. We previously argued that there are no quintic quotients \eqref{Z5Z5quinticQ} with $\IZ_3$ or $\IZ_5$ symmetries. Additionally we find that there are no quintic quotients with a $\text{SL}(2,\IZ_3)$ symmetry. Consider the generators of this group $g_i$ with corresponding co-ordinate actions $\pi_i$ and induced linear actions $A_{\pi_i}$ on the coefficients $c_a$. It is possible to show by direct computation that the intersection of eigenspaces of  $A_{\pi_i}$ and $A_{\pi_j}$ for $i\ne j$ is trivial. Thus there does not exist any non-null $c\in \IC^6$ for which the polynomial \eqref{Z5Z5quinticQ} is preserved, up to a multiple. Since there are no quotients with
$\text{SL}(2,\IZ_3)$ symmetry, there cannot be any with $\text{SL}(2,\IZ_5){\;\supset\;}\text{SL}(2,\IZ_3)$.
\vskip30pt
\section{Galois actions}\label{sec:Galois}
\vskip-10pt
We have been careful not to specify a value for $\z$, beyond the fact that it is a nontrivial fifth root of unity. Now we enquire as to how our calculations change if we replace $\z$ by $\z^\k$, with $\k\in \{1,2,3,4\}$. 

Consider the effect of this transformation, which we shall refer to as a $\k$-transformation, on the normalizer condition \eqref{QuinticNormalizerCond1}. In this
relation the matrix $T$ is independent of $\z$, while the matrix $S$ does depend on $\z$.
\beq
S(\z)\;=\;\text{diag}(1,\,\z,\z^2,\z^3,\z^4)~~~\text{and so}~~~S(\z^\k)\;=\;S(\z)^\k~.
\notag\eeq
From this observation it is easy to see that if $\p(\z)$ satisfies the normalizer condition with matrix 
$\s{\;=\;}\left(\begin{smallmatrix} \alpha & \beta \\\gamma & \delta \end{smallmatrix}\right)$ then $\p(\z^\k)$
satisfies an analogous relation with matrix
\beq
{}^\k \s \;=\; 
\left(\begin{matrix} \a & \k^{-1}\b \\ \k\g & \d \end{matrix}\right)~,
\label{ktransform}\eeq
where $\k^{-1}$ denotes the inverse in $\IZ^{\text{*}}_5$. 
Thus a $\k$-transform has the effect
\beq
(\s,\,\p(\z),\,A(\z)) \rightarrow ({}^\k \s,\,\p(\z^\k),\, A(\z^\k))~,
\notag\eeq
which we may think of as induced by an action of the Galois group on $\z$, or more strictly, on the roots of the irreducible polynomial, of which $\z$ is a root 
\beq
1+\z +\z^2 +\z^3 +\z^4 \;=\; 0~.
\notag\eeq
This observation simplifies our task significantly, since the $\s$-matrices fall into orbits $\{ {}^\k \s \}$ as 
$\k$ varies. Consider, again, the case of families with $\IZ_4$ symmetry. We learn from GAP, or from \tref{SL_2_5_Subgroups}, that $\text{SL}(2,\IZ_5)$ has 15 distinct $\IZ^4$ subgroups. These fall into
3 orbits of length 4, one orbit of length 2 and one orbit of length 1, as $\k$ varies over the Galois group. Since we understand the action of $\k$ on the eigenvectors of $A$ we need only calculate these for one $\s$ taken from each orbit, so for 5 matrices rather than 15.

The orbits we are counting, in this example, are orbits of distinct $\IZ_4$ subgroups. `Short' orbits arise if, for example, the matrix $\s$ is invariant under $\k$-transformation, as is the case for 
\beq
\s\;=\;\left(\begin{matrix} 3 & 0 \\ 0 & 2 \end{matrix}\right)~.
\notag\eeq
It can also happen that the matrices ${}^\k \s$, as $\k$ varies, are distinct but they do not generate 4 distinct 
$\IZ_4$ subgroups. An example is provided by
\beq
\s\;=\;\left(\begin{matrix} 0 & 1 \\ 4 & 0 \end{matrix}\right)~,
\notag\eeq
for which ${}^4\s{\,=\,}({}^1\s)^{-1}$ and ${}^3\s{\,=\,}({}^2\s)^{-1}$, so only two $\IZ_4$ groups are generated.

We should explain a general point that turns out to relate particularly to the families with precisely $\IZ_6$ symmetry of \tref{TZ6}. We have seen that we have to find the eigenvectors of matrices $A_\pi$ with elements in the field 
$\IQ(\z)$. The first step in finding such eigenvectors is to solve the characteristic polynomial for the eigenvalues.
The characteristic polynomial has coefficients in $\IQ(\z)$. In general, we would expect this polynomial to factor over some higher field. Somewhat surprisingly, in all cases, apart from the cases corresponding to symmetry precisely 
$\IZ_6$, the eigenvalues take values in $\IQ(\z)$ itself. For $\IZ_6$, however, we find that the eigenvalues take values in $\IQ(\z,\o)$, with $\o$ a nontrivial cube root of unity. An elementary application of the primitive element theorem ensures that we can achieve this field with a single extension. In this case $\IQ(\z,\o) \,=\, \IQ(\e)$ where 
$\e\,=\,\o\z$. This quantity is a fifteenth root of unity, but it is not a general such root. There are four choices for $\z$ and two for $\o$, so eight choices for $\e$. Indeed $\e$ satisfies the following irreducible polynomial equation of degree~eight
\beq
1-\epsilon+\epsilon ^3\!-\epsilon ^4+\epsilon ^5\!-\epsilon ^7+\epsilon ^8 \;=\; 0~.
\label{cyclotomic}\eeq
Given $\z$, there are two choices for $\e$. These are the common roots of the above polynomial with the relation
$\e^6{\,=\,}\z$. The polynomial above is $\Phi_{15}(\e)$, the fifteenth cyclotomic polynomial, and the roots are 
$\e^k$ for $k$ coprime to 15, that is for the eight values 
\beq
\cG\;\defineas\;\{1,\,2,\,4,\,7,\,8,11,13,14\}~.
\notag\eeq 
The two common roots of \eqref{cyclotomic} and the equation $\e^6{\,=\,}\z$ are $\e$ and $\e^{11}$.
This quantity $\e$ is important to us because it is the quantity that appears in \tref{TZ6}.

Now, the set $\cG$ is in fact a group, with the group operation being multiplication mod 15. Since $\cG$ permutes the roots of \eqref{cyclotomic} via
\beq
\e^k \mapsto \e^{jk}~;~~~j,\,k\in \cG~,
\notag\eeq
it is the Galois group of the polynomial. As an abstract group, 
$\cG{\;\cong\;}\IZ_2{\,\times\,}\IZ_4$; in terms of generators, we can write $\cG{\;=\;}\langle 11,\, 2\rangle$.

For the case of families that have precisely $\IZ_6$ symmetry, each $\s$-matrix gives rise to two curves of symmetric manifolds. The group $\cG$ has the order two element that maps $\e{\,\mapsto\,} \e^{11}$, which we have noted leaves $\z$ invariant. In each case, this interchanges the two curves. The order 4 element that maps 
$\e{\,\mapsto\,} \e^2$ induces a $\k$-transform via the relation $\z{\,=\,}\e^6$, and this acts as~before. 
\vskip30pt
\section{Smooth and singular manifolds}\label{sec:smooth}
\vskip-10pt
The generic quintic is smooth and so is its generic $\IZ_5{\times}\IZ_5$ quotient. However, upon further restriction of the defining polynomial, in order to satisfy the requirements of enhanced symmetry, we are led to polynomials which are far from being generic. We therefore have to check whether the symmetric manifolds are in fact smooth. 

To check whether a manifold, defined by the equation $p(x){\,=\,}0$, is singular, we have to check whether the five equations $\del p/\del x_i{\,=\,}0$ have a simultaneous solution. Let us take again a $\IZ_4$ family as illustrative. 

Corresponding to the second row of \tref{TZ4}, we have
\beq
p(x) \;=\; \a_1 \mathbf{J_1} + \a_2(\mathbf{J_2} + \z^{4}\mathbf{J_3}) + 
\a_3(\mathbf{J_4} + \z^{3}\mathbf{J_5}) + \mathbf{J_6}~.
\notag\eeq
We want to know if the conditions are satisfied, for generic values of the coefficients $\a_1,\,\a_2,\,\a_3$. The quickest method seems to be to assign `random' integer values to the $\a$'s and to solve the equations numerically. This is done first with $x_0{\,=\,}1$. Then we set $x_0{\,=\,}0$ and $x_1{\,=\,}1$ and solve again, then we set
$x_0{\,=\,}x_1{\,=\,}0$ and $x_2{\,=\,}1$, and so on. In this way we find all solutions, if they exist. For generic values of the coefficients we find there are no solutions, so we conclude that the $\IZ_4$ manifold above is smooth. The same holds for the remaining families of $\IZ_4$ manifolds. We could, in principle, avoid the need for numerical computation by performing a Gr\"obner basis calculation, but these calculations can be slow and do not always complete.

For the $\IQ_8$ manifolds, the analogous calculation yields four solutions, for generic values of the parameters. For each of these solutions we compute the determinant of the $4{\,\times\,}4$ matrix 
\beq
\frac{\del^2 p(x)}{\del x_i \del x_j}~,
\notag\eeq
on the relevant patches. We find that the determinants are nonzero, so we conclude that these singularities are conifolds. We proceed analogously for the remaining cases. 

To summarize: the generic quotients, with $\IZ_4$, Dic$_3$ and Dic$_5$ symmetries, are smooth. However the generic manifolds of the $\IZ_6$ families have 4 conifold singularities\footnote{By this we mean 4 conifold points on the quotient. On the $\IP^4$ there are, of course, more singular points, which are then identified under the action of $S$ and $T$.}, as do the manifolds of the $\IQ_8$ families. These conifold points organise themselves into two orbits of length two under the corresponding $\IZ_6$ or $\IQ_8$ group action.  

The $\IZ_{10}$ manifolds have no parameters, and each such manifold is singular, having, at least, a singular curve. This is a case where the Gr\"obner basis calculation seems the best way to proceed.
\vskip30pt
\section{Tables}\label{results_quintic_quotients}
\vskip-10pt
We have noted in \sref{SymmQuinQuo} that each symmetric polynomial corresponds to a left eigenspace of a matrix
$A_\pi$. Such an eigenspace corresponds to a family of symmetric manifolds, and has a basis of eigenvectors, the general eigenvector being a linear combination of these. In the tables, the basis eigenvectors become polynomials. Basis polynomials, for a given eigenspace, are separated by commas. The generic polynomial is then a linear combination of these. Distinct eigenspaces correspond to distinct families of manifolds. These are separated, in the tables, by vertical space. Thus the first row of \tref{TZ4}, for example, corresponds to a single, three-parameter, family of manifolds, while the first row of \tref{TQ8} corresponds to three, one-parameter, families of manifolds. 

The $\IZ_2$ generator $\sigma=-\one_2$ generating the symmetry action $\pi^{(2)}$ defined in \eqref{eq:quintic_new_symm1} is omitted from the tables with $\IZ_6$ and $\IZ_{10}$ symmetries, since $\pi^{(2)}$ is a  symmetry of all the quotients. Instead, the generators for the  $\IZ_3$ and $\IZ_{5}$ groups are presented respectively for brevity.
\subsection{$\IZ_4$ Families}
\begin{table}[H]
\vspace{0pt}
\footnotesize
\begin{center}
\begin{tabular}{|c | c | c | c|}
\hline
\varstr{15pt}{10pt} \textbf{\small{$\sigma$}}  & \textbf{Symm Action} & \textbf{$\IZ_{4}$ Covariants} & \# \textbf{Fams.} \\ \hline				
\hline 
$\left(
\begin{array}{cc}
4 & 4 \\
2 & 1 \\
\end{array}
\right)$ & ${\left(
\begin{array}{ >{\hskip-3pt} c <{\hskip-3pt} c <{\hskip-3pt} c <{\hskip-3pt} c <{\hskip-3pt} c <{\hskip-3pt} }
 1 & \zeta  & \zeta  & 1 & \zeta ^3 \\
 1 & \zeta ^2 & \zeta ^3 & \zeta ^3 & \zeta ^2 \\
 \zeta  & \zeta ^4 & \zeta  & \zeta ^2 & \zeta ^2 \\
 \zeta ^3 & \zeta ^2 & 1 & \zeta ^2 & \zeta ^3 \\
 \zeta  & \zeta  & 1 & \zeta ^3 & 1 \\
\end{array}
\right)}$ & \footnotesize\begin{minipage}[c][85pt][c]{2.5in}
\begin{gather*}
\mathbf{J_2} -\z^4\mathbf{J_3},\\
\mathbf{J_4 }- \z^3 \mathbf{J_5},\\
\mathbf{J_1} +(\z^2+\z^4)\mathbf{J_2} +(\z+\z^2)\mathbf{J_5},\\
4\mathbf{J_1}+10(\z^2+\z^4)\mathbf{J_2} +\mathbf{J_6}\\
 \end{gather*}
\end{minipage} & 4 \\ 				
\hline 
$\left(
\begin{array}{cc}
2 & 0 \\
1 & 3 \\
\end{array}
\right)$ & $
{\left(
\begin{array}{ >{\hskip-3pt} c <{\hskip-3pt} c <{\hskip-3pt} c <{\hskip-3pt} c <{\hskip-3pt} c <{\hskip-3pt} }
 1 & 0 & 0 & 0 & 0 \\
 0 & 0 & \zeta ^4 & 0 & 0 \\
 0 & 0 & 0 & 0 & 1 \\
 0 & \zeta ^3 & 0 & 0 & 0 \\
 0 & 0 & 0 & \zeta ^3 & 0 \\
\end{array}
\right)}$  & \footnotesize\begin{minipage}[c][85pt][c]{2.5in}
\begin{gather*}
\mathbf{J_1},\\
\mathbf{J_2}+\z^{4}\mathbf{J_3},\\
\mathbf{J_4}+\z^{3}\mathbf{J_5},\\
\mathbf{J_6} \\
 \end{gather*}
\end{minipage} & 4 \\ 
\hline 
$\left(
\begin{array}{cc}
2 & 1 \\
0 & 3 \\
\end{array}
\right)$ & $
{\left(
\begin{array}{ >{\hskip-3pt} c <{\hskip-3pt} c <{\hskip-3pt} c <{\hskip-3pt} c <{\hskip-3pt} c <{\hskip-3pt} }
 1 & \zeta  & 1 & \zeta ^2 & \zeta ^2 \\
 \zeta ^2 & \zeta ^2 & 1 & \zeta  & 1 \\
 \zeta  & 1 & \zeta ^2 & \zeta ^2 & 1 \\
 \zeta ^2 & 1 & \zeta  & 1 & \zeta ^2 \\
 1 & \zeta ^2 & \zeta ^2 & 1 & \zeta  \\
\end{array}
\right)}$  & \footnotesize\begin{minipage}[c][85pt][c]{2.5in}
\begin{gather*}
\mathbf{J_2} -\z^3\mathbf{J_3},\\
\mathbf{J_4} -\z\mathbf{J_5},\\
\mathbf{J_1} +(1+\z^2)\mathbf{J_2} +(1+\z)\mathbf{J_5},\\
4\mathbf{J_1} +10(1+\z^2)\mathbf{J_2} +\mathbf{J_6}\\
\end{gather*}
\end{minipage} & 4 \\ 
\hline 
 {$\left(\begin{array}{cc}
0 & 1 \\
4 & 0 \\
\end{array}
\right)$} & \footnotesize
${\left(
\begin{array}{ >{\hskip-3pt} c <{\hskip-3pt} c <{\hskip-3pt} c <{\hskip-3pt} c <{\hskip-3pt} c <{\hskip-3pt} }
 1 & 1 & 1 & 1 & 1 \\
 1 & \zeta ^4 & \zeta ^3 & \zeta ^2 & \zeta  \\
 1 & \zeta ^3 & \zeta  & \zeta ^4 & \zeta ^2 \\
 1 & \zeta ^2 & \zeta ^4 & \zeta  & \zeta ^3 \\
 1 & \zeta  & \zeta ^2 & \zeta ^3 & \zeta ^4 \\
\end{array}
\right)}$
& \footnotesize\begin{minipage}[c][85pt][c]{2.5in}
\begin{gather*}
 \mathbf{J_2}-\mathbf{J_3},\\
 \mathbf{J_4}-\mathbf{J_5},\\
\mathbf{J_1}+(\zeta +\zeta^4)\mathbf{J_2}+(\zeta^2+\zeta^3)\mathbf{J_5},\\
4\mathbf{J_1}+10(\zeta +\zeta^4)\mathbf{J_2}+\mathbf{J_6}\\
\end{gather*}
\end{minipage} & 2  \\ 	
\hline 
$\left(
\begin{array}{cc}
3 & 0 \\
0 & 2 \\
\end{array}
\right)$ & ${\left(
\begin{array}{ >{\hskip-3pt} c <{\hskip-3pt} c <{\hskip-3pt} c <{\hskip-3pt} c <{\hskip-3pt} c <{\hskip-3pt} }
 1 & 0 & 0 & 0 & 0 \\
 0 & 0 & 0 & 1 & 0 \\
 0 & 1 & 0 & 0 & 0 \\
 0 & 0 & 0 & 0 & 1 \\
 0 & 0 & 1 & 0 & 0 \\
\end{array}
\right)}$  & \footnotesize\begin{minipage}[c][85pt][c]{2.5in}
\begin{gather*}
\mathbf{J_1},\\\mathbf{J_2}+\mathbf{J_3},\\\mathbf{J_4}+\mathbf{J_5},\\\mathbf{J_6} \\
\end{gather*}
\end{minipage} & 1  \\ 
\hline
\end{tabular}
\vskip20pt
\capt{5.8in}{TZ4}{The families of manifolds with $\IZ_{4}$ symmetries. Each of the $\sigma$ matrices of the first three rows generate four $\IZ_4$ subgroups of $\text{SL}(2,\IZ_5)$ under the $\kappa$-transformation \eqref{ktransform}. Each of these rows give rise, in this way, to 4 distinct three-parameter families of manifolds. The penultimate 
$\sigma$ matrix forms an orbit of length 2 under $\k$-transformation and so gives rise to 2 distinct three-parameter families of manifolds. The $\sigma$ matrix of the final row is fixed by $\k$-transformation, and so gives rise to 1 three-parameter family of manifolds.}
\end{center}
\end{table}
\newpage
\subsection{$\IZ_6$  Families}
\begin{table}[H]
\vspace{-5pt}
\footnotesize
\begin{center}
\begin{tabular}{| >{\hskip-3pt} c <{\hskip-3pt} | >{\hskip-12pt} c <{\hskip-12pt} |c|}
\hline
\varstr{15pt}{10pt}  \textbf{\small{$\sigma$}}  & \textbf{Symm Action} & \textbf{$\IZ_6$ Covariants} \\ \hline
\hline 
$\left(
\begin{array}{ >{\hskip-3pt} c <{\hskip-3pt} c <{\hskip-3pt} }
4 & 1 \\
4 & 0 \\
\end{array}
\right)$ 
& 
$\left(
\begin{array}{ >{\hskip-3pt} c <{\hskip-3pt} c <{\hskip-3pt} c <{\hskip-3pt} c <{\hskip-3pt} c <{\hskip-3pt} }
 1 & 1 & 1 & 1 & 1 \\
 \zeta ^4 & \zeta ^3 & \zeta ^2 & \zeta  & 1 \\
 \zeta ^2 & 1 & \zeta ^3 & \zeta  & \zeta ^4 \\
 \zeta ^4 & \zeta  & \zeta ^3 & 1 & \zeta ^2 \\
 1 & \zeta  & \zeta ^2 & \zeta ^3 & \zeta ^4 \\
\end{array}
\right)$
& \footnotesize\begin{minipage}[c][170pt][c]{310pt}
\vspace{20pt}
{\begin{align*}
&   4 \mathbf{J_1}{-}5
   \left(2{-}\epsilon{-}\epsilon ^2{+}\epsilon ^3{-}2 \epsilon ^4{+}\epsilon ^5{-}2
   \epsilon ^7\right) \mathbf{J_2}{-}
   5 (1{-}\epsilon )^2 (1{+}\epsilon ) \mathbf{J_3}{+}\mathbf{J_6}\,, \\[5pt]
& 2 \left({-}1+2 \epsilon +\epsilon ^2-\epsilon ^3+\epsilon ^4{-}\epsilon ^5+2
   \epsilon ^7\right) \mathbf{J_1}{-}\!
   \left(2+\epsilon+\epsilon^3+\epsilon^5+2 \epsilon^6\right) \mathbf{J_2}\\
&\hskip10pt-\left(2-2 \epsilon ^2+\epsilon ^5+\epsilon ^6-\epsilon^7\right) \mathbf{J_3}\\
&\hskip10pt   +\left(5-3 \epsilon -4 \epsilon ^2+3 \epsilon ^3-2 \epsilon ^4+4   \epsilon ^5+3 \epsilon ^6-5 \epsilon ^7\right) \mathbf{J_4}+\mathbf{J_5}\\[15pt]
&   4 \mathbf{J_1}{+}5 \left(1{-}\epsilon
   {+}\epsilon ^3{-}2 \epsilon ^4{+}\epsilon ^5{+}\epsilon ^6{-}\epsilon ^7\right) \mathbf{J_2}{-}
   5   \left(1{+}\epsilon {+}\epsilon ^3{+}\epsilon ^6{-}\epsilon ^7\right) \mathbf{J_3}{+}\mathbf{J_6}\,, \\[5pt]
& 2 \left(1-2\epsilon +\epsilon ^2-\epsilon ^4+\epsilon ^5-\epsilon ^6\right)
   \mathbf{J_1}-\left(1-\epsilon +\epsilon ^3-\epsilon ^5+\epsilon ^6\right)\mathbf{J_2}\\
&\hskip10pt  -\left(1-\epsilon ^2-\epsilon ^5+\epsilon ^6-2 \epsilon ^7\right)\mathbf{J_3}\\
&\hskip10pt   -\left(1-3 \epsilon +3 \epsilon ^2-\epsilon ^3-2 \epsilon ^4+4 \epsilon^5-4 \epsilon ^6+2 \epsilon ^7\right) \mathbf{J_4}+\mathbf{J_5}\\
\end{align*}}
\end{minipage} \\ 									
\hline 
$\left(
\begin{array}{ >{\hskip-3pt} c <{\hskip-3pt} c <{\hskip-3pt} }
 1 & 1 \\
2 & 3 \\
\end{array}
\right)$ & 
${\left(
\begin{array}{ >{\hskip-3pt} c <{\hskip-3pt} c <{\hskip-3pt} c <{\hskip-3pt} c <{\hskip-3pt} c <{\hskip-3pt} }
 1 & \zeta ^3 & \zeta ^4 & \zeta ^3 & 1 \\
 \zeta  & \zeta ^3 & \zeta ^3 & \zeta  & \zeta ^2 \\
 \zeta ^3 & \zeta ^4 & \zeta ^3 & 1 & 1 \\
 \zeta  & \zeta  & \zeta ^4 & 1 & \zeta ^4 \\
 1 & \zeta ^4 & \zeta  & \zeta  & \zeta ^4 \\
\end{array}
\right)}$
& \footnotesize\begin{minipage}[c][173pt][c]{310pt}
\vspace{20pt}
{\begin{align*}
& 4 \mathbf{J_1}-5
   \left(\epsilon ^2+\epsilon ^4+\epsilon ^5+\epsilon ^7\right) \mathbf{J_2}-5
   \left(\epsilon -\epsilon ^3+\epsilon ^5\right) \mathbf{J_3}+\mathbf{J_6}\,,\\[5pt]
& 2 \left(2-\epsilon -\epsilon ^2+\epsilon ^3-\epsilon ^4+2 \epsilon^5+\epsilon ^6-2 \epsilon ^7\right) \mathbf{J_1}\\
&\hskip10pt +\left(3-\epsilon -2 \epsilon^2+\epsilon ^3-2 \epsilon ^4+\epsilon ^5+\epsilon ^6-3 \epsilon ^7\right)
   \mathbf{J_2}\\
&\hskip10pt   -\left(1-2 \epsilon -\epsilon ^2+2 \epsilon ^3-\epsilon ^4+\epsilon
   ^5+\epsilon ^6-2 \epsilon ^7\right) \mathbf{J_3}\\
&\hskip10pt   +\left(3+\epsilon -2 \epsilon^2-\epsilon ^3-2 \epsilon ^4+\epsilon ^5+3 \epsilon ^6\right) \mathbf{J_4}+\mathbf{J_5}  \\[15pt]
& 4 \mathbf{J_1}-5(1-\epsilon )^2 \epsilon ^3 (1+\epsilon ) \mathbf{J_2}+5 \left(1+\epsilon +\epsilon^3+\epsilon ^5+\epsilon ^6\right) \mathbf{J_3}+\mathbf{J_6}\,,\\[5pt]
& 2 \left(-1+\epsilon-\epsilon ^2+\epsilon ^4-2 \epsilon ^5+\epsilon ^6\right) \mathbf{J_1}+
   \epsilon \left(1-\epsilon -\epsilon ^2+2 \epsilon ^3-\epsilon ^4\right) \mathbf{J_2}\\
&\hskip10pt +\left(1-2\epsilon +\epsilon ^2-\epsilon ^3-\epsilon ^4+\epsilon ^5-2 \epsilon ^6\right)
   \mathbf{J_3}-\epsilon  \left(1-\epsilon +\epsilon ^2\right)^2 \mathbf{J_4}+\mathbf{J_5}\\
   \end{align*}}
\end{minipage}  \\ 				
\hline 
  $\left(
\begin{array}{ >{\hskip-3pt} c <{\hskip-3pt} c <{\hskip-3pt} }
 2 & 1 \\
3 & 2 \\
\end{array}
\right)$ & \begin{minipage}[c][85pt][c]{1.6in}
\begin{gather*}
{\left(
\begin{array}{ >{\hskip-3pt} c <{\hskip-3pt} c <{\hskip-3pt} c <{\hskip-3pt} c <{\hskip-3pt} c <{\hskip-3pt} }
 1 & \zeta ^2 & \zeta  & \zeta ^2 & 1 \\
 \zeta ^2 & \zeta ^3 & \zeta  & \zeta  & \zeta ^3 \\
 \zeta  & \zeta  & \zeta ^3 & \zeta ^2 & \zeta ^3 \\
 \zeta ^2 & \zeta  & \zeta ^2 & 1 & 1 \\
 1 & \zeta ^3 & \zeta ^3 & 1 & \zeta ^4 \\
\end{array}
\right)}\\
\end{gather*}
\end{minipage}  & \footnotesize\begin{minipage}[c][153pt][c]{310pt}
\vspace{20pt}
{\begin{align*}
& 4 \mathbf{J_1}+5 \left(2{-}\epsilon {+}\epsilon
   ^3{-}\epsilon ^4{+}\epsilon ^5{-}2 \epsilon ^7\right) \mathbf{J_2}+5 \epsilon 
   \left(1{-}\epsilon {+}\epsilon ^3{-}\epsilon ^4{+}\epsilon ^6\right) \mathbf{J_3}+\mathbf{J_6}\,,\\[5pt]
& 2 \left(1+\epsilon -\epsilon ^4+\epsilon ^6\right) \mathbf{J_1}+\left(2-\epsilon
   -\epsilon ^2+2 \epsilon ^3-\epsilon ^4+\epsilon ^5-3 \epsilon ^7\right)
   \mathbf{J_2}\\
&\hskip10pt +\epsilon  \left(1{+}\epsilon {-}\epsilon ^2{-}\epsilon ^3{+}\epsilon ^5{+}\epsilon
   ^6\right) \mathbf{J_3}{+}
   \left(2+3 \epsilon {-}\epsilon ^3{-}\epsilon ^4-\epsilon ^5{+}2
   \epsilon ^6+2 \epsilon ^7\right) \mathbf{J_4}{-}\mathbf{J_5} \\[15pt]
& 4 \mathbf{J_1}{+}5 \epsilon  \left(1{-}\epsilon{ +}\epsilon ^3{-}\epsilon ^4{+}\epsilon^6\right) \mathbf{J_2}{+}5 \left(2{-}\epsilon{+}\epsilon ^3{-}\epsilon ^4{+}\epsilon ^5{-}2
   \epsilon ^7\right) \mathbf{J_3}+\mathbf{J_6}\,,\\[5pt]
& 2\epsilon  \left(1-\epsilon +\epsilon ^2-\epsilon ^3+\epsilon ^5-\epsilon^6\right) \mathbf{J_1}-
   \epsilon  \left(1-2 \epsilon +\epsilon ^2+\epsilon ^3-\epsilon^4\right) \mathbf{J_2}\\
&\hskip10pt   +\left(1+\epsilon -2 \epsilon ^2+2 \epsilon ^3-\epsilon
   ^4+\epsilon ^6-2 \epsilon ^7\right) \mathbf{J_3}\\
&\hskip10pt   +\left(-2+3 \epsilon -3 \epsilon ^2+2\epsilon ^3-\epsilon ^4-\epsilon ^5+2 \epsilon ^6-\epsilon ^7\right)\mathbf{J_4}+\mathbf{J_5}\\
   \end{align*}}
\end{minipage}  \\ 
\hline
\end{tabular}
\vskip10pt
\capt{6.2in}{TZ6}{The families of $\IZ_{6}$ manifolds. Together with $\s=-\one_2$, each of the $\sigma$ matrices of the first two rows generate four $\IZ_6$ subgroups under $\kappa$-transformation, and so give rise to 8 distinct one-parameter families each. The final row of $\s$ matrix along with $\s=-\one_2$, generates two $\IZ_6$ subgroups, under 
$\k$-transformation and gives rise to four $\IZ_6$ families.}
\end{center}
\end{table}
\subsection{The $\IQ_8$ Families}\label{Q8families}
\begin{table}[H]
\vspace{0pt}
\footnotesize
\begin{center}
\begin{tabular}{|c | c | c | >{\hskip-3pt} c <{\hskip-3pt} |}
\hline
\varstr{15pt}{10pt} \textbf{\small{$\sigma$}}  & \textbf{Symm Action} & \textbf{$\IQ_{8}$ Covariants} & \# \textbf{Fams.} \\ \hline		
\hline  
\begin{minipage}[c][160pt][c]{0.5in}
\begin{gather*}
{\left(\begin{matrix}
 2 & 0 \\
 1 & 3 \\
\end{matrix}\right)}\\[46pt]
{\left(\begin{matrix}
 1 & 1 \\
 3 & 4 \\
\end{matrix}\right)}\\
\end{gather*}
\end{minipage}
 & \begin{minipage}[c][85pt][c]{1.6in}
\begin{gather*}
{\left(
\begin{array}{ >{\hskip-3pt} c <{\hskip-3pt} c <{\hskip-3pt} c <{\hskip-3pt} c <{\hskip-3pt} c <{\hskip-3pt} }
 1 & 0 & 0 & 0 & 0 \\
 0 & 0 & \zeta ^4 & 0 & 0 \\
 0 & 0 & 0 & 0 & 1 \\
 0 & \zeta ^3 & 0 & 0 & 0 \\
 0 & 0 & 0 & \zeta ^3 & 0 \\
\end{array}
\right)}\\[8pt]
{\left(
\begin{array}{ >{\hskip-3pt} c <{\hskip-3pt} c <{\hskip-3pt} c <{\hskip-3pt} c <{\hskip-3pt} c <{\hskip-3pt} }
 1 & \zeta  & \zeta  & 1 & \zeta ^3 \\
 \zeta  & \zeta  & 1 & \zeta ^3 & 1 \\
 \zeta ^3 & \zeta ^2 & 1 & \zeta ^2 & \zeta ^3 \\
 \zeta  & \zeta ^4 & \zeta  & \zeta ^2 & \zeta ^2 \\
 1 & \zeta ^2 & \zeta ^3 & \zeta ^3 & \zeta ^2 \\
\end{array}
\right)}\\
\end{gather*}
\end{minipage}  & \begin{minipage}[c][170pt][c]{3in}
{\begin{gather*}
 \mathbf{J_2}-\zeta^4  \mathbf{J_3}, \\\mathbf{J_4}- \zeta ^3 \mathbf{J_5}\\[15pt]
  4 \mathbf{J_1}+5 \left(1+\zeta\right) \mathbf{J_2}+5 \left(1+\zeta ^4\right) \mathbf{J_3}+\mathbf{J_6},\\
 6  \mathbf{J_1}+5  \left(1+\zeta^2\right) \mathbf{J_4}+5 \left(1+\zeta ^3\right) \mathbf{J_5}-\mathbf{J_6} \\[15pt]
4 \mathbf{J_1}+5 \left(\zeta^2+\zeta ^4\right) \mathbf{J_2}+ 5 \left(\zeta +\zeta^3\right) \mathbf{J_3}+\mathbf{J_6},\\
6 \mathbf{J_1} +5\left(\zeta^3+\zeta^4\right) \mathbf{J_4} +5\left(\zeta+\zeta^2\right) \mathbf{J_5}-\mathbf{J_6}
\end{gather*}}
\end{minipage} & 12\\ 
\hline 
\begin{minipage}[c][160pt][c]{0.5in}
\begin{gather*}
{\left(\begin{matrix}
 3 & 0 \\
 0 & 2 \\
\end{matrix}\right)}\\[46pt]
{\left(\begin{matrix}
 0 & 4 \\
 1 & 0 \\
\end{matrix}\right)}\\
\end{gather*}
\end{minipage}
 & \begin{minipage}[c][85pt][c]{1.6in}
\begin{gather*}
{\left(\begin{matrix}
1 & 0 & 0 & 0 & 0 \\
 0 & 0 & 0 & 1 & 0 \\
 0 & 1 & 0 & 0 & 0 \\
 0 & 0 & 0 & 0 & 1 \\
 0 & 0 & 1 & 0 & 0 \\
\end{matrix}\right)}\\[8pt]
{\left(
\begin{array}{ >{\hskip-3pt} c <{\hskip-3pt} c <{\hskip-3pt} c <{\hskip-3pt} c <{\hskip-3pt} c <{\hskip-3pt} }
 1 & 1 & 1 & 1 & 1 \\
 1 & \zeta  & \zeta ^2 & \zeta ^3 & \zeta ^4 \\
 1 & \zeta ^2 & \zeta ^4 & \zeta  & \zeta ^3 \\
 1 & \zeta ^3 & \zeta  & \zeta ^4 & \zeta ^2 \\
 1 & \zeta ^4 & \zeta ^3 & \zeta ^2 & \zeta  \\
\end{array}
\right)}\\
\end{gather*}
\end{minipage}  & \begin{minipage}[c][170pt][c]{3in}
{\begin{gather*}
 \mathbf{J_2}-\mathbf{J_3}, \\\mathbf{J_4}-\mathbf{J_5} \\[15pt]
4 \mathbf{J_1} +5 \left(\zeta ^2+\zeta^3\right) (\mathbf{J_2} +\mathbf{J_3} )+\mathbf{J_6},\\
6 \mathbf{J_1} +5 \left(\zeta +\zeta^4\right) (\mathbf{J_4}+\mathbf{J_5})- \mathbf{J_6} \\[15pt]
 4 \mathbf{J_1}+5 \left(\zeta +\zeta ^4\right) (\mathbf{J_2}+\mathbf{J_3})+\mathbf{J_6},\\
6 \mathbf{J_1}+5 \left(\zeta ^2+\zeta ^3\right) (\mathbf{J_4}+\mathbf{J_5})- \mathbf{J_6} \\
\end{gather*}}
\end{minipage} & 3\\ 
\hline
\end{tabular}
\vskip20pt
\capt{6.0in}{TQ8}{Families of manifolds with $\IQ_{8}$ symmetry. Each row corresponds, initially, to 3 
one-parameter families. The first row gives rise to an orbit of length 4 under $\k$-transformation, so to a total of  twelve one-parameter $\IQ_8$ families. The second row is invariant under $\k$-transformation, so gives rise to 3 such families.} 
\end{center}
\end{table}
\newpage
\subsection{The rigid $\IZ_{10}$ manifolds}
\begin{table}[H]
\vspace{0pt}
\footnotesize
\begin{center}
\begin{tabular}{|c |  >{\hskip-3pt} c <{\hskip-3pt}  | >{\hskip-10pt} c <{\hskip-10pt} |  >{\hskip-3pt} c <{\hskip-3pt}  |}
\hline
\varstr{15pt}{10pt} \textbf{\small{$\sigma$}}  & \textbf{Symm Action} & \textbf{$\IZ_{10}$ Covariants} & \# \textbf{Quotients} \\ \hline
\hline  
$\left(
\begin{array}{cc}
 0 & 1 \\
4 & 2 \\
\end{array}
\right)$ &
${\left(
\begin{array}{ >{\hskip-3pt} c <{\hskip-3pt} c <{\hskip-3pt} c <{\hskip-3pt} c <{\hskip-3pt} c <{\hskip-3pt} }
 1 & 1 & \zeta ^2 & \zeta  & \zeta ^2 \\
 1 & \zeta ^4 & 1 & \zeta ^3 & \zeta ^3 \\
 1 & \zeta ^3 & \zeta ^3 & 1 & \zeta ^4 \\
 1 & \zeta ^2 & \zeta  & \zeta ^2 & 1 \\
 1 & \zeta  & \zeta ^4 & \zeta ^4 & \zeta  \\
\end{array}
\right)}$  & \begin{minipage}[c][95pt][c]{3.2in}
\begin{gather*}
4 \mathbf{J_1}+5 \left(1+\zeta^4\right) \mathbf{J_2}+5 \left(\zeta^2 +\zeta ^4\right) \mathbf{J_3}+\mathbf{J_6}\\[3pt]
4 \mathbf{J_1}+5  \left(\zeta+\zeta ^3\right) \mathbf{J_2}+5 \left(1+\zeta \right)\mathbf{J_3}+\mathbf{J_6} \\[3pt]
6 \mathbf{J_1}+5 \left(\zeta +\zeta^2\right) \mathbf{J_4}+5 \left(1+\zeta ^2\right) \mathbf{J_5}-\mathbf{J_6} \\[3pt]
6 \mathbf{J_1}+5 \left(1+\zeta ^3\right) \mathbf{J_4}+5 \left(\zeta ^3+\zeta^4\right) \mathbf{J_5}-\mathbf{J_6}\\
 \end{gather*}
\end{minipage} & 16  \\ 				
\hline  
 $\left(
\begin{array}{cc}
 1 & 0 \\
4 & 1 \\
\end{array}
\right)$ & $
{\left(
\begin{array}{ >{\hskip-3pt} c <{\hskip-3pt} c <{\hskip-3pt} c <{\hskip-3pt} c <{\hskip-3pt} c <{\hskip-3pt} }
 1 & 0 & 0 & 0 & 0 \\
 0 & \zeta ^4 & 0 & 0 & 0 \\
 0 & 0 & \zeta ^2 & 0 & 0 \\
 0 & 0 & 0 & \zeta ^4 & 0 \\
 0 & 0 & 0 & 0 & 1 \\
\end{array}
\right)}$  & \footnotesize\begin{minipage}[c][95pt][c]{3.2in}
\begin{gather*}
 \mathbf{J_2} \\[3pt]
 \mathbf{J_3} \\[3pt]
 \mathbf{J_4} \\[3pt]
 \mathbf{J_5} \\
\end{gather*} 
\end{minipage} & 4 \\ 			
\hline
$\left(
\begin{array}{cc}
 1 & 1 \\
0 & 1 \\
\end{array}
\right)$ & $
{\left(
\begin{array}{ >{\hskip-3pt} c <{\hskip-3pt} c <{\hskip-3pt} c <{\hskip-3pt} c <{\hskip-3pt} c <{\hskip-3pt} }
 1 & 1 & \zeta  & \zeta ^3 & \zeta  \\
 \zeta  & 1 & 1 & \zeta  & \zeta ^3 \\
 \zeta ^3 & \zeta  & 1 & 1 & \zeta  \\
 \zeta  & \zeta ^3 & \zeta  & 1 & 1 \\
 1 & \zeta  & \zeta ^3 & \zeta  & 1 \\
\end{array}
\right)}$  & \footnotesize\begin{minipage}[c][95pt][c]{3.2in}
\begin{gather*}
4 \mathbf{J_1}+5 \left(\z^2+\z^3\right) \mathbf{J_2} +5 \left(\zeta+\zeta ^4\right) \mathbf{J_3}+\mathbf{J_6} \\[3pt]
4 \mathbf{J_1} +5 \left(\zeta +\zeta ^4\right) \mathbf{J_2}+5 \left(\zeta ^2+\zeta^3\right)  \mathbf{J_3}+\mathbf{J_6} \\[3pt]
6 \mathbf{J_1}+ 5 \left(\zeta ^2+\zeta^3\right)  \mathbf{J_4}+5 \left(\zeta +\zeta ^4\right) \mathbf{J_5}-\mathbf{J_6} \\[3pt]
6 \mathbf{J_1}+5 \left(\zeta +\zeta ^4\right) \mathbf{J_4}+ 5 \left(\zeta ^2+\zeta^3\right)  \mathbf{J_5}-\mathbf{J_6} \\
 \end{gather*}
\end{minipage} & 4 \\			
\hline
\end{tabular}
\vskip20pt
\capt{6.0in}{TZ10}{Rigid manifolds with $\IZ_{10}$ symmetry. Each row corresponds, initially, to four rigid 
$\IZ_{10}$ manifolds. The first row has an orbit of length four under $\k$-transformation so gives rise to a total of sixteen manifolds. The second and third rows are invariant under $\k$-transformation so give rise to 4 manifolds each.}
\end{center}
\end{table}
\newpage
\subsection{$\text{Dic}_3$ Families}
\begin{table}[H]
\vspace{5pt}
\footnotesize
\begin{center}
\begin{tabular}{|c |  >{\hskip-3pt} c <{\hskip-3pt}  | c |>{\hskip-4pt}c<{\hskip-4pt}|}
\hline
\varstr{15pt}{10pt} \textbf{\small{$\sigma$}}  & \textbf{Symm Action} & \textbf{Dic$_{3}$ Covariants} & \# \textbf{Fams.}\\ \hline
\hline
\begin{minipage}[c][160pt][c]{0.5in}
\begin{gather*}
{\left(\begin{matrix}
 3 & 4 \\
 0 & 2 \\
\end{matrix}\right)}\\[46pt]
{\left(\begin{matrix}
 0 & 3 \\
 3 & 4 \\
\end{matrix}\right)}\\
\end{gather*}
\end{minipage}
 & \begin{minipage}[c][150pt][c]{1.4in}
\begin{gather*}
{\left(
\begin{array}{ >{\hskip-3pt} c <{\hskip-3pt} c <{\hskip-3pt} c <{\hskip-3pt} c <{\hskip-3pt} c <{\hskip-3pt} }
 1 & \zeta  & 1 & \zeta ^2 & \zeta ^2 \\
 1 & \zeta ^2 & \zeta ^2 & 1 & \zeta  \\
 \zeta ^2 & 1 & \zeta  & 1 & \zeta ^2 \\
 \zeta  & 1 & \zeta ^2 & \zeta ^2 & 1 \\
 \zeta ^2 & \zeta ^2 & 1 & \zeta  & 1 \\
\end{array}
\right)}\\[8pt]
{\left(
\begin{array}{ >{\hskip-3pt} c <{\hskip-3pt} c <{\hskip-3pt} c <{\hskip-3pt} c <{\hskip-3pt} c <{\hskip-3pt} }
 1 & 1 & \zeta ^3 & \zeta ^4 & \zeta ^3 \\
 1 & \zeta ^3 & \zeta ^4 & \zeta ^3 & 1 \\
 1 & \zeta  & 1 & \zeta ^2 & \zeta ^2 \\
 1 & \zeta ^4 & \zeta  & \zeta  & \zeta ^4 \\
 1 & \zeta ^2 & \zeta ^2 & 1 & \zeta  \\
\end{array}
\right)}\\
\end{gather*}
\end{minipage}  & \begin{minipage}[c][150pt][c]{3.2in}
\vspace{30pt}
{\begin{align*}
& 4 \mathbf{J_1}{-}10 \left(1+\z+2 \zeta ^3\right) \mathbf{J_2}{+}10 \zeta \left(1-\z+\zeta ^2\right) \mathbf{J_3}+\mathbf{J_6}\,,\\[5pt]
& \mathbf{J_1}-\left(1+ \zeta+2 \zeta^3\right) \mathbf{J_2} +\zeta \left(1-\zeta+ \zeta ^2\right) \mathbf{J_3}\\
&\hskip10pt -(1+\zeta)^2\mathbf{J_4} +(\z+\z^2-\zeta^4)  \mathbf{J_5}\\
 \end{align*}}
\end{minipage} & 4\\ 		
\hline
 \begin{minipage}[c][160pt][c]{0.5in}
\begin{gather*}
{\left(\begin{matrix}
 3 & 0 \\
 1 & 2 \\
\end{matrix}\right)}\\[46pt]
{\left(\begin{matrix}
 1 & 1 \\
 2 & 3 \\
\end{matrix}\right)}\\
\end{gather*}
\end{minipage}
 & \begin{minipage}[c][150pt][c]{1.4in}
\begin{gather*}
{\left(
\begin{array}{ >{\hskip-3pt} c <{\hskip-3pt} c <{\hskip-3pt} c <{\hskip-3pt} c <{\hskip-3pt} c <{\hskip-3pt} }
 1 & 0 & 0 & 0 & 0 \\
 0 & 0 & 0 & \zeta ^2 & 0 \\
 0 & \zeta ^2 & 0 & 0 & 0 \\
 0 & 0 & 0 & 0 & 1 \\
 0 & 0 & \zeta  & 0 & 0 \\
\end{array}
\right)}\\[8pt]
{\left(
\begin{array}{ >{\hskip-3pt} c <{\hskip-3pt} c <{\hskip-3pt} c <{\hskip-3pt} c <{\hskip-3pt} c <{\hskip-3pt} }
 1 & \zeta ^3 & \zeta ^4 & \zeta ^3 & 1 \\
 \zeta  & \zeta ^3 & \zeta ^3 & \zeta  & \zeta ^2 \\
 \zeta ^3 & \zeta ^4 & \zeta ^3 & 1 & 1 \\
 \zeta  & \zeta  & \zeta ^4 & 1 & \zeta ^4 \\
 1 & \zeta ^4 & \zeta  & \zeta  & \zeta ^4 \\
\end{array}
\right)}\\
\end{gather*}
\end{minipage}  & \begin{minipage}[c][150pt][c]{3.2in}
\vspace{30pt}
{\begin{align*}
& 4 \mathbf{J_1} -10 \zeta(1+\zeta)^2  \mathbf{J_2} + 10 \left(1+\z-\z^3\right) \mathbf{J_3}+\mathbf{J_6},\\[5pt] 
& \mathbf{J_1}-\zeta  (1+\zeta )^2 \mathbf{J_2}+\left(1+\zeta -\zeta ^3\right) \mathbf{J_3}\\
&\hskip10pt + \left(1+\z^3-\z^4\right)\mathbf{J_4} +\left(1-\zeta +\zeta ^2\right) \mathbf{J_5}\\
\end{align*}}
\end{minipage} & 4\\ 	
\hline
\begin{minipage}[c][160pt][c]{0.5in}
\begin{gather*}
{\left(\begin{matrix}
 2 & 0 \\
 0 & 3 \\
\end{matrix}\right)}\\[46pt]
{\left(\begin{matrix}
 2 & 2 \\
 4 & 2 \\
\end{matrix}\right)}\\
\end{gather*}
\end{minipage}
 & \begin{minipage}[c][150pt][c]{1.4in}
\begin{gather*}
{\left(\begin{matrix}
1 & 0 & 0 & 0 & 0 \\
 0 & 0 & 1 & 0 & 0 \\
 0 & 0 & 0 & 0 & 1 \\
 0 & 1 & 0 & 0 & 0 \\
 0 & 0 & 0 & 1 & 0 \\
 \end{matrix}\right)}\\[8pt]
{\left(
\begin{array}{ >{\hskip-3pt} c <{\hskip-3pt} c <{\hskip-3pt} c <{\hskip-3pt} c <{\hskip-3pt} c <{\hskip-3pt} }
 1 & 1 & \zeta  & \zeta ^3 & \zeta  \\
 \zeta ^4 & \zeta  & \zeta ^4 & \zeta ^3 & \zeta ^3 \\
 \zeta ^4 & \zeta ^3 & \zeta ^3 & \zeta ^4 & \zeta  \\
 1 & \zeta  & \zeta ^3 & \zeta  & 1 \\
 \zeta ^2 & 1 & \zeta ^4 & \zeta ^4 & 1 \\
\end{array}
\right)}\\
\end{gather*}
\end{minipage}  & \begin{minipage}[c][150pt][c]{3.2in}
\vspace{30pt}
{\begin{align*}
& 4 \mathbf{J_1}-10 \left(2+\zeta ^2+\zeta ^3\right) (\mathbf{J_2}+\mathbf{J_3})+\mathbf{J_6}\,,\\[5pt]
& 6 (\mathbf{J_2}+ \mathbf{J_3})-4 \left(2-3 \zeta ^2-3 \zeta ^3\right) (\mathbf{J_4}+\mathbf{J_5})\\
&\hskip10pt -\left(1-\zeta ^2-\zeta ^3\right) \mathbf{J_6}\\
\end{align*}}
\end{minipage} & 2\\
\hline
\end{tabular}
\vskip15pt
\capt{5.8in}{TDic3}{The families with $\text{Dic}_{3}$ symmetry. Each row corresponds, initially to one-parameter families. The first two rows admit orbits of length 4 under $\k$-transformation, so each give rise to 4 one-parameter families. The third row admits an orbit of length 2 under $\k$-transformation, so to two distinct Dic$_3$ families.}
\end{center}
\end{table}
\newpage
\subsection{$\text{Dic}_5$ Families}
\begin{table}[H]
\vspace{10pt}
\footnotesize
\begin{center}
\begin{tabular}{|c | c | c |c|}
\hline
\varstr{15pt}{10pt} \textbf{\small{$\sigma$}}  & \textbf{Symm Action} & \textbf{Dic$_{5}$ Covariants} & \# \textbf{Fams.} \\ \hline
\hline
 \begin{minipage}[c][160pt][c]{0.5in}
\begin{gather*}
{\left(\begin{matrix}
 0 & 1 \\
 4 & 0 \\
\end{matrix}\right)}\\[46pt]
{\left(\begin{matrix}
 1 & 3 \\
 1 & 4 \\
\end{matrix}\right)}\\
\end{gather*}
\end{minipage}
 & \begin{minipage}[c][145pt][c]{1.6in}
\begin{gather*}
{\left(
\begin{array}{ >{\hskip-3pt} c <{\hskip-3pt} c <{\hskip-3pt} c <{\hskip-3pt} c <{\hskip-3pt} c <{\hskip-3pt} }
 1 & 1 & 1 & 1 & 1 \\
 1 & \zeta ^4 & \zeta ^3 & \zeta ^2 & \zeta  \\
 1 & \zeta ^3 & \zeta  & \zeta ^4 & \zeta ^2 \\
 1 & \zeta ^2 & \zeta ^4 & \zeta  & \zeta ^3 \\
 1 & \zeta  & \zeta ^2 & \zeta ^3 & \zeta ^4 \\
\end{array}
\right)}\\[8pt]
{\left(
\begin{array}{ >{\hskip-3pt} c <{\hskip-3pt} c <{\hskip-3pt} c <{\hskip-3pt} c <{\hskip-3pt} c <{\hskip-3pt} }
 1 & \zeta ^4 & \zeta  & \zeta  & \zeta ^4 \\
 \zeta ^4 & \zeta  & \zeta  & \zeta ^4 & 1 \\
 1 & 1 & \zeta ^3 & \zeta ^4 & \zeta ^3 \\
 \zeta ^3 & \zeta  & \zeta ^2 & \zeta  & \zeta ^3 \\
 \zeta ^3 & \zeta ^4 & \zeta ^3 & 1 & 1 \\
\end{array}
\right)}\\
\end{gather*}
\end{minipage}  & \begin{minipage}[c][145pt][c]{2.5in}
\vspace{10pt}
{\begin{gather*}
4 \mathbf{J_1}+10 \z^4{\mathbf{J_2}}+ 10\zeta \mathbf{J_3}+\mathbf{J_6},\\[2pt]
6 \mathbf{J_1}+10\zeta^3 \mathbf{J_4}+10 \zeta^2 {\mathbf{J_5}}-\mathbf{J_6}\\
\end{gather*}}
\end{minipage} & 4\\ 					
\hline 
\begin{minipage}[c][160pt][c]{0.5in}
\begin{gather*}
{\left(\begin{matrix}
 3 & 0 \\
 4 & 2 \\
\end{matrix}\right)}\\[46pt]
{\left(\begin{matrix}
 3 & 0 \\
 1 & 2 \\
\end{matrix}\right)}\\
\end{gather*}
\end{minipage}
 & \begin{minipage}[c][145pt][c]{1.6in}
\begin{gather*}
{\left(\begin{matrix}
1 & 0 & 0 & 0 & 0 \\
 0 & 0 & 0 & {\zeta^3} & 0 \\
 0 & {\zeta^3} & 0 & 0 & 0 \\
 0 & 0 & 0 & 0 & 1 \\
 0 & 0 & {\zeta^4} & 0 & 0 \\
 \end{matrix}\right)}\\[8pt]
{\left(
\begin{array}{ >{\hskip-3pt} c <{\hskip-3pt} c <{\hskip-3pt} c <{\hskip-3pt} c <{\hskip-3pt} c <{\hskip-3pt} }
 1 & 0 & 0 & 0 & 0 \\
 0 & 0 & 0 & \zeta ^2 & 0 \\
 0 & \zeta ^2 & 0 & 0 & 0 \\
 0 & 0 & 0 & 0 & 1 \\
 0 & 0 & \zeta  & 0 & 0 \\
\end{array}
\right)}\\
\end{gather*}
\end{minipage}  & \begin{minipage}[c][145pt][c]{2.5in}
\vspace{20pt}
{\begin{gather*}
\mathbf{J_1}\,,\\
\mathbf{J_6}~ \\
\end{gather*}}
\end{minipage} & 1\\ 	
\hline
\begin{minipage}[c][160pt][c]{0.5in}
\begin{gather*}
{\left(\begin{matrix}
 2 & 2 \\
 0 & 3 \\
\end{matrix}\right)}\\[46pt]
{\left(\begin{matrix}
 2 & 0 \\
 0 & 3 \\
\end{matrix}\right)}\\
\end{gather*}
\end{minipage}
 & \begin{minipage}[c][145pt][c]{1.6in}
\begin{gather*}
{\left(
\begin{array}{ >{\hskip-3pt} c <{\hskip-3pt} c <{\hskip-3pt} c <{\hskip-3pt} c <{\hskip-3pt} c <{\hskip-3pt} }
 1 & \zeta ^4 & \zeta ^2 & \zeta ^4 & 1 \\
 \zeta ^4 & 1 & 1 & \zeta ^4 & \zeta ^2 \\
 \zeta ^4 & \zeta ^2 & \zeta ^4 & 1 & 1 \\
 1 & 1 & \zeta ^4 & \zeta ^2 & \zeta ^4 \\
 \zeta ^2 & \zeta ^4 & 1 & 1 & \zeta ^4 \\
\end{array}
\right)}\\[8pt]
{\left(\begin{matrix}
 1 & 0 & 0 & 0 & 0 \\
 0 & 0 & 1 & 0 & 0 \\
 0 & 0 & 0 & 0 & 1 \\
 0 & 1 & 0 & 0 & 0 \\
 0 & 0 & 0 & 1 & 0 \\
\end{matrix}\right)}\\
\end{gather*}
\end{minipage}  & \begin{minipage}[c][145pt][c]{2.5in}
\vspace{20pt}
{\begin{gather*}
6 \mathbf{\mathbf{J_1}}+10 ({\mathbf{J_4}}+ {\mathbf{J_5}})-\mathbf{\mathbf{J_6}}, \\[2pt]
4 \mathbf{J_1} +10 (\mathbf{J_2}+\mathbf{J_3})+\mathbf{J_6}\\
\end{gather*}}
\end{minipage} & 1\\
\hline
\end{tabular}
\vskip15pt
\capt{6.0in}{TDic5}{The families with $\text{Dic}_{5}$ symmetry. Each row corresponds, initially, to a one-parameter family. The first row admits an orbit of length 4 under $\k$-transformation. The second and third rows are invariant under $\k$-transformation, so give rise to one family each.}
\end{center}
\end{table}

\section*{Acknowledgments}
\vskip-10pt
The authors would like to thank Xenia de la Ossa and Andre Lukas for helpful discussions. PC is supported by EPSRC grant BKRWDM00 and wishes to acknowledge the hospitality of KIAS, where part of this work was done. 
\vskip3cm
\renewcommand{\baselinestretch}{0.9}\normalsize
\bibliographystyle{utcaps}
\bibliography{bibfile}

\providecommand{\href}[2]{#2}\begingroup\raggedright\begin{thebibliography}{10}

\bibitem{Braun:2010vc}
V.~Braun, ``{On Free Quotients of Complete Intersection Calabi-Yau
  Manifolds},'' {\em JHEP} {\bf 1104} (2011) 005,
\href{http://arXiv.org/abs/1003.3235}{{\tt 1003.3235}}.

\bibitem{Candelas:2008wb}
P.~Candelas and R.~Davies, ``{New Calabi-Yau Manifolds with Small Hodge
  Numbers},'' {\em Fortsch.Phys.} {\bf 58} (2010) 383--466,
\href{http://arXiv.org/abs/0809.4681}{{\tt 0809.4681}}.

\bibitem{Candelas:2010ve}
P.~Candelas and A.~Constantin, ``{Completing the Web of $Z_3$ - Quotients of
  Complete Intersection Calabi-Yau Manifolds},'' {\em Fortsch.Phys.} {\bf 60}
  (2012) 345--369,
\href{http://arXiv.org/abs/1010.1878}{{\tt 1010.1878}}.

\bibitem{candelas2016hodge}
P.~Candelas, A.~Constantin, and C.~Mishra, ``Hodge numbers for CICYs with
  symmetries of order divisible by 4,'' {\em Fortschritte der Physik} {\bf 64}
  (2016), no.~6-7, 463--509.

\bibitem{constantin2017hodge}
A.~Constantin, J.~Gray, and A.~Lukas, ``Hodge numbers for all CICY quotients,''
  {\em Journal of High Energy Physics} {\bf 2017} (2017), no.~1, 1.

\bibitem{candelas2016calabi}
P.~Candelas, A.~Constantin, and C.~Mishra, ``Calabi-Yau Threefolds With Small
  Hodge Numbers,'' {\em arXiv:1602.06303} (2016).

\bibitem{Rodland:1998pm}
E.~A. R{\nordico}dland, ``The Pfaffian Calabi--Yau, its mirror, and their link
  to the Grassmannian G(2, 7),'' {\em Compositio Mathematica} {\bf 122} (2000),
  no.~02, 135--149.

\bibitem{freitag2011siegel}
E.~Freitag and R.~S. Manni, ``On Siegel three folds with a projective
  Calabi--Yau model,'' {\em arXiv:1103.2040} (2011).

\bibitem{Witten:1985xc}
E.~Witten, ``{Symmetry Breaking Patterns in Superstring Models},'' {\em
  Nucl.Phys.} {\bf B258} (1985)
75.

\bibitem{goodman1986global}
M.~W. Goodman and E.~Witten, ``Global symmetries in four and higher
  dimensions,'' {\em Nuclear Physics B} {\bf 271} (1986), no.~3-4, 21--52.

\bibitem{luhn2008quintics}
C.~Luhn and P.~Ramond, ``Quintics with finite simple symmetries,'' {\em Journal
  of Mathematical Physics} {\bf 49} (2008), no.~5, 053525.

\bibitem{conway1985atlas}
J.~H. Conway, R.~T. Curtis, S.~P. Norton, and R.~A. Parker, ``Atlas of finite
  groups,''.

\bibitem{GAP4}
The GAP~Group, {\em {GAP -- Groups, Algorithms, and Programming, Version
  4.8.7}}, 2017.

\end{thebibliography}\endgroup
\end{document}